\documentclass[aps,prc,floatfix,amsmath]{revtex4}
\usepackage{dcolumn}
\usepackage{amssymb}
\usepackage{mathrsfs}
\usepackage{textcomp}
\usepackage{bm}
\usepackage{supertabular}
\usepackage{color,graphicx}
\usepackage[bookmarksopen]{hyperref}
\newcommand{\bea}{\begin{eqnarray}}
\newcommand{\eea}{\end{eqnarray}}
\newcommand{\nn}{\nonumber\\}

\begin{document}
\title{Non-Fermi liquid behavior of thermal relaxation time in degenerate electron plasma}
\author{Sreemoyee Sarkar}
\email{sreemoyee.sarkar@saha.ac.in}

\author{Abhee K. Dutt-Mazumder}
\email{abhee.dm@saha.ac.in}

\affiliation{High Energy Nuclear and Particle Physics Division, Saha Institute of Nuclear Physics,
1/AF Bidhannagar, Kolkata-700 064, INDIA}

\medskip

\begin{abstract}
The thermal relaxation time ($\tau_{\kappa_{ee}}$) for the degenerate electron plasma has been calculated by incorporating non-Fermi liquid (NFL)
corrections
 both for the thermal conductivity and specific heat capacity. Perturbative results are presented by making expansion
 in $T/m_D$ with next to leading order corrections (NLO). We see that the NLO NFL corrections further reduce the decrease 
in relaxation time due to the leading order (LO) correction.
 
%  We see that the NLO NFL corrections partially offset the decrease 
% in relaxation time due to the leading order (LO) correction.
% {\bf It is seen that unlike the normal Fermi liquid (FL) result, NFL corrections in leading order (LO) reduces
%  the thermal relaxation time.} Incorporation of the phase space correction driven by the medium modified Fermion dispersion 
% relation increases the relaxation time further. 

\end{abstract}

\maketitle
\section{Introduction}
Determination of thermal relaxation time of degenerate electron matter has been a subject of
serious investigation for the last several decades. Application of such studies encompass
broad areas spreading across various disciplines like metals, semiconductors, astrophysical
objects like white dwarfs or neutron stars to name a few. Our focus here is to study the
heat conduction in neutron star. In particular, we determine the thermal relaxation time
of degenerate electron system at high density which is relevant for heat transfer from
the crust of a star to the core\cite{Pethick92, Lattimer94, Gnedin01}. 

It is known that when a new star is born following a supernova explosion, large amount of
neutrinos are emitted immediately from the core resulting in colder core and a hotter
crust, thus a temperature gradient is set up. Then the thermal energy gradually flows inward by heat conduction 
which alternatively might be viewed as the propagation of the cooling waves from the 
center towards the surface leading to thermalization \cite{Gnedin01}. One of the subjects of contemporary
research in astrophysics has been the estimation of this thermalization time scale or estimation
of the thermal relaxation time. The investigation what we pursue here is relevant in the context of neutron star
as we know that degenerate electron gas and positively charged ion constitute the envelop of
the crust of the neutron stars.  

There exists several calculations where heat conduction has been studied extensively. In these
investigations major contributions have been seen to originate from the electron-ion scattering \cite{Potekhin99, Gnedin01}. 
The
contribution of the electron-electron scattering, in contrast, have been found to be of limited importance \cite{Potekhin99}. 
Recently this problem has been revisited and it was seen that such conclusions are true only
when one considers the charge-charge interaction and neglects the current-current interaction
completely. This is a valid approximation in dealing with the ions, but might not be  justified
for the electron-electron scattering where at high density the magnetic interaction becomes 
important \cite{Heiselberg93, Shternin06, Shternin07}. Reference \cite{Shternin06}, in particular, deals with the calculation of 
heat conductivity where
it has been shown that at high density, due to strong magnetic interaction, the electron-electron collision frequency
become larger than the electron-ion collision frequency reducing the heat conductivity.
The other point is to note that almost all these calculations treat the degenerate electron matter as ideal Fermi liquid and treat the electron-electron
 and electron-ion scattering non-relativistically restricting to the electric sector. But at high density for the electrons
 with momentum close to the Fermi momentum, since relativistic effects become important, the magnetic interaction can no longer
 be neglected. It is now known that with the inclusion of the transverse interaction the normal Fermi liquid description
 breaks down due to the vanishing of the electron propagator near the Fermi surface. This can be attributed to the absence
 of static screening of the magnetic photon \cite{Gerhold05}. Several investigations have been performed
 in recent years where incorporation of such corrections have been seen to have serious implications on various physical
 quantities like pressure, entropy, viscosity or quantities like drag and diffusion coefficients \cite{Ipp04, Gerhold204, Nandi, Sarkar11}. 
Non-Fermi liquid behavior for the neutrino emissivity or the 
neutrino mean free path have
also been studied extensively \cite{Schafer04, Pal11,Adhya12}. In all these calculations such NFL corrections have been observed
 to be significant compared to the Fermi liquid results.

In this work, therefore we incorporate NFL corrections while estimating the thermal relaxation
 time ($\tau_{\kappa_e}$) of degenerate electron plasma. Such estimation requires knowledge of both the thermal conductivity ($\kappa_{e}$)
 and specific heat ($c_v$) where this correction has to be included consistently for a given order. 
Furthermore in dense matter the quasiparticle dispersion characteristics
 change which modify the density of states too. Inclusion of this, as we shall see, also modifies the results significantly both for
  $\kappa_e$ and $\tau_{\kappa_e}$. To derive analytical expressions for these quantities however, we make 
perturbative expansion in $T/m_D$, where, $T$ is the plasma temperature and $m_D$ is the Debye mass.

The plan of the paper is as follows. In section II we develop the formalism part which incorporates the Boltzmann equation, 
the screening mechanism of long-range interactions and the evaluation of the thermal relaxation time. In subsection A the results of leading order thermal conductivity and the thermal relaxation time have
 been discussed and in the subsection B next to leading order NFL correction of the thermal conductivity and the 
thermal relaxation time have been included followed by summary and conclusion.

%%%%%%%%%%%%%%%%%%%%%%%%%%%%%%%%%%%%%%%%%%%%%%%%%%%%%%%%%%%%%%%%%%%%%%%%%
\section{Formalism}
% \subsection{Medium Modified Phase Space Factor}
%%%%%%%%%%%%%%%%%%%%%%%%%%%%%%%%%%%%%%%%%%%%%%%%%%%%%%%%%%%%%%%%%%%%%%%%%
% In this section we 
We consider degenerate electron gas where the electrons constitute an almost ideal and uniform Fermi gas and collide between 
 themselves. We aim to calculate the thermal relaxation time of degenerate electron gas with the medium modified 
phase space factor. The characteristic relaxation time for thermal conduction 
$\tau_{\kappa}$ can be defined as follows \cite{Heiselberg93},
\bea
\tau_{\kappa}=\frac{3\kappa}{c_v}.
\label{relax_time}
\eea
In case of the strongly degenerate 
electron gas the electron thermal conductivity ($\kappa_e$) can be expressed as follows,
\bea
\kappa_e=\frac{ J_T}{T\nu_{e}}, \quad \nu_e=\nu_{ei}+\nu_{ee},
\label{def_k}
\eea
where, $J_T$ is the thermal current and $\nu_e$ is the total effective collision frequency. This one is the sum of the partial
 collision frequencies, {\em i.e}, electron-ion ($\nu_{ei}$) and electron-electron ($\nu_{ee}$) scattering rate. Evidently $\kappa_e$ is related to
 $\kappa_{ee}$ and $\kappa_{ei}$ \cite{Shternin06},
\bea
 && { 1 \over \kappa_e} ={ 1 \over \kappa_{ei}} +{ 1 \over \kappa_{ee}},
  \nonumber \\
 &&  
   \kappa_{ei}=\frac{J_T}{T\nu_{ei}},\quad\kappa_{ee}=\frac{J_T}{T\nu_{ee}}.
\eea
Hence, derivation of $\kappa_{ee}$ in turn requires the knowledge of $\nu_{ee}$. It is clear from the denominator of the above equation that the heat conduction becomes difficult when collision frequency increases.
 
To evaluate $\kappa_{ee}$ we appeal to the Boltzmann equation which describes the kinetics for the individual fermion
component \cite{Heiselberg93}, 
\bea
\left(\frac{\partial}{\partial t}+{\bf v_p}.\nabla_{\bf r} +{\bf F}.\nabla_{{\bf p}}\right)f_p=-\mathcal{C}[f_p],
\label{boltz_tot}
\eea 
here, ${\bf p}$ is the momentum of the quasiparticle, {\bf F} is the external force, ${\bf v_p}$ is the velocity of the heat carrier
and $f_p$ is the distribution function of electrons. The collision-integral on the right-hand-side (RHS) is given by the rate of fermions scattering in and out of the state with 
momentum ${\bf p}$ by scattering on the other fermions with momentum ${\bf p^{'}}$.
In the presence of a weak stationary
 temperature gradient and absence of any external force the Boltzmann equation takes the following form,
\bea
{\bf v_{p}}.\nabla_{\bf r} f_p=-\mathcal{C}[f_p].
\label{boltz}
\eea 
 Now, due to presence of a weak 
temperature gradient these Fermi-Dirac distribution functions deviate from equilibrium distribution functions $f_i$, which we write
 as,
\bea
    \widetilde{f_i}=f_i +    
    {\partial f_i  \over \partial \epsilon_i}\Phi_i\,\frac{\nabla T}{T}
\label{distributions}
\eea 
where,
\bea 
    f_i= \left\{ \exp \left( {\epsilon_i - \mu \over
    T} \right) +1 \right\}^{-1}.
\eea

$\epsilon$ is the particle energy, $\mu$ is the chemical potential and $T$ is the temperature. Clearly the second
 term with $\Phi$ measures the deviation
 from equilibrium. The collision integral can be written as follows,
\bea
\mathcal{C}[f_p]=\nu \nu^{'}\int_{p^{'},k,k^{'}}\widetilde{ f_p}\widetilde{f_k} (1\pm\widetilde{ f_p^{'}})(1\pm 
\widetilde{f_k^{'}})-\widetilde{f_p^{'}}\widetilde{f_k^{'}}(1\pm \widetilde{f_p})(1\pm \widetilde{f_k})(2\pi)^4 
\delta^4(p+p^{'}-k-k^{'})| M | ^2.
\label{collisionint}
\eea
% $\nu$, $\nu^{'}$ are the degeneracy factors. 
$|M|^2$ is the squared matrix element for the scattering process $12\rightarrow 34$. 
The $\pm$ sign include stimulated emission and Pauli blocking. In this paper, we only consider the electron-electron scattering, hence,
 from now onwards only negative sign will be considered in the phase space factor. Using the standard linearization procedure from Eqs.(\ref{boltz}) and (\ref{distributions}) we obtain equation for $\Phi$, 
\bea
f_p(1-f_p)(\epsilon_p -\mu)v_{pz}=\nu \nu^{'} \int_{p^{'},k,k^{'}} 
  f_pf_k (1- f_p^{'})(1-f_k^{'})(2\pi)^4 \delta^4(p+k-p^{'}-k^{'})| M | ^2 (\Phi_p+\Phi_k-\Phi_{p^{'}}-\Phi_{k^{'}}).
\label{therm_cond}
\eea
The above equation can be written in the form $|X\rangle=I |\Phi\rangle$,
 where, $|X\rangle=(\epsilon_p -\mu)v_p$ and $I$ is the integral operator. 
% the right-hand-side of the Eq.(\ref{therm_cond}) which
% acts on $\Phi$.
 The thermal conductivity $\kappa_{ee}$ is given by the maximum of the following equation and has already been discussed in 
\cite{Heiselberg93, Baym},
\bea
\kappa_{ee}=\frac{\langle X|\Phi\rangle^2}{T\langle \Phi|I|\Phi\rangle},
\label{kappa}
\eea
 $\langle\cdot |\cdot\rangle$ denotes an inner product, the quantity 
$\frac{\langle X|\Psi\rangle^2}{T\langle \Psi|I|\Psi\rangle}$ is minimal for $\Psi=\Phi$ with the minimal value $\kappa_{ee}$.
This is another way to define $\kappa_{ee}$ in Eq.(\ref{def_k}). Hence, one can write,
\bea
\frac{1}{\kappa_{ee}}&\geq& \left(\nu\int_p \frac{(\epsilon_p -\mu)}{T}v_zf_p(1-f_p)\Psi_p\right)^{-2}\nn\
&\times&\nu \nu^{'}\int_{p,p^{'},k,k^{'}} 
  f_pf_k (1- f_p^{'})(1- f_k^{'})(2\pi)^4 \delta^4(p+k-p^{'}-k^{'})| M | ^2 \frac{(\Psi_p+\Psi_k
-\Psi_{p^{'}}-\Psi_{k^{'}})^2}{4},
\label{def_cond}
\eea
the term in the first bracket in the denominator is the thermal current $J_T$. $\Phi$ should be determined by minimizing Eq.(\ref{kappa}) and the minimal value is $\Psi$. But for the present purpose here we consider
 the simplest trial function \cite{Heiselberg93, Shternin06},
\bea
\Psi_p\propto (\epsilon_p -\mu)v_z.
\eea
The above trial function can now be inserted in Eq.(\ref{def_cond}) and the term in the bracket can be averaged 
over the $z$ axis keeping
 $x$ and $\phi$ fixed, where, $\phi$ is the azimuthal angle between $v_p$ and $v_k$. After averaging we obtain \cite{Heiselberg93},
 \bea
(\Psi_p+\Psi_k-\Psi_{p^{'}}-\Psi_{k^{'}})^2=\frac{2}{3}\omega^2(1-x^2)(1-\text{cos}\phi).
\eea
To proceed further one needs to know the interaction. Considering only the electron-electron scattering the 
squared matrix element for small energy transfer is given by \cite{Heiselberg93},
% \bea
% \lvert M \rvert ^2 =32 e^4\frac{s^2+u^2}{t^2}.
% \lvert M \rvert ^2 =e^4\frac{u^2+s^2}{t^2}.
% \label{matamp_qq}
% \eea
% The above matrix amplitude squared is infrared divergent in the $t\rightarrow 0$ limit. To circumvent this problem one needs 
% to consider the medium modified screening effect.
% {\bf We in this paper use the Braaten and Pisarski's prescription 
% \cite{Braaten90}. In this method in the matrix amplitude squared the hard-dense-loop (HDL) resummed propagator \cite{Manuel96, Manuel97} 
% is incorporated in the lower momentum region and in the higher momentum region one can use the bare propagator. 
% On addition the arbitrary momentum scale gets canceled.
%  Using the in medium HDL resummed propagator the matrix amplitude squared for the electron electron interaction takes the following form,}
\bea
| M|^2 =32e^4\left[\frac{1}{\left(q^2+\Pi_L\right)}+
\frac{(1-x^2)\mbox{cos}\phi}{(q^2-\omega^2+\Pi_T)}\right]^2
\label{mat_amp_exp}.
\eea
In the above equation the medium modified photon propagator contains the polarization functions $\Pi_L (q,\omega)$ and $\Pi_T (q,\omega)$, which describe 
plasma screening
of interparticle interaction by longitudinal and transverse
plasma perturbations, respectively. 
For small momentum transfers ($q\ll \mu$) \cite{Manuel96, Manuel97},
% are the longitudinal and the transverse part of the dressed photon propagator and in the HDL framework the structure of these two polarization tensors are the following,

% which depend on ! and q and describe .
\bea
\Pi_L (q,\omega)=m_D^2\chi_L,\quad \Pi_T (q,\omega)=m_D^2\chi_T,
\label{pol_tensor}
\eea
where,
\bea
\chi_L&=&\left[1-\frac{x}{2}\mbox{ln}\left(\frac{x+1}{x-1}\right) \right],\nn\
\chi_T&=&\left[\frac{x^2}{2}+\frac{x(1-x^2)}{4}\mbox{ln}\left(\frac{x+1}{x-1}\right) \right].
\eea
In the above expressions $x=q_0/q$ and $m_D$ is the Debye mass $m_D^2=e^2\mu^2/\pi^2$.  For $q_0^2<q^2$ the logarithmic term has 
an imaginary contribution,
\bea
\mbox{ln}\left(\frac{q_0+q}{q_0-q}\right)=\mbox{ln}\left|\frac{q_0+q}{q_0-q}\right|-i\pi\theta\left(q^2-q_0^2\right).
\eea
In case of the thermal conductivity due to presence of $\text{cos}\phi$ term in the numerator of Eq.(\ref{def_cond}), 
the cross term of the matrix amplitude squared does not vanish, hence, on retaining the cross term the matrix amplitude
 squared becomes, 
\bea
| M|^2 =32e^4\left[\frac{1}{\left(q^2+m_D\right)^2}-\frac{2 q^4 \text{cos}\phi}{\left(q^2+m_D^2\right)\left(q^6+
\frac{\pi^2\omega^2m_D^4}{16}\right)}+
\frac{q^2\mbox{cos}^2\phi}{q^6+\frac{\pi^2\omega^2m_D^4}{16}}\right].
\eea

% With this the Eq.(\ref{def_cond}) can be rewritten as,
Now, we first compute the denominator of Eq.(\ref{def_cond}). The denominator is the thermal current $J_T$ as indicated earlier and is given by,
\bea
J_T&=&\nu\int_p \frac{(\epsilon_p -\mu)}{T}v_zf_p(1-f_p)\Psi_p\nn\
&=& \frac{\nu \mu^2 T^2}{6},
% &=&\frac{\mu^2T^2 }{3}.
\eea
 where we use the following equation, 
\bea
\int_{-\infty}^{\infty}\frac{x^2}{e^x+1}\left(1-\frac{1}{e^x+1}\right)dx=\frac{\pi^2}{3}.
\eea
% The above equation is true both for the electrons and the quarks. 
Now, for electrons the degeneracy factor is $\nu=2$
and $J_T$ for the electrons is then $\mu^2T^2/3$. On the other hand in case of quarks for each flavor ($\nu=6$), 
the thermal current becomes
 $\mu^2T^2$.
% In case of quark-quark scattering the thermal current takes the following form,
% \bea
% J_T&=& \frac{\nu \mu^2 T^2}{6}\nn\
% &=&\mu^2T^2.
% \eea

To proceed further to evaluate the numerator in Eq.(\ref{def_cond}) it is convenient to introduce a dummy integration 
variable $\omega$. We write the energy conserving delta function as,
\bea
\delta(\epsilon_p+\epsilon_k-\epsilon_p^{'}-\epsilon_k^{'})=\int_{-\infty}^{\infty}d\omega 
\delta(\omega+\epsilon_p-\epsilon_p^{'})\delta(\omega-\epsilon_k+\epsilon_k^{'}).
\eea
Evaluating ${\bf q}={\bf p^{'}}-{\bf p}$ in terms of p, q and $\mbox{cos} \theta_{pq}$ and defining $t=\omega^2-q^2$ we find,
\bea
\delta(\omega+\epsilon_p-\epsilon_p^{'})&&=\frac{p^{'}}{×pq}\delta\left(\mbox{cos} \theta_{pq}-\frac{\omega}{×q}-\frac{t}{2pq×}\right)\nn\
\delta(\omega-\epsilon_k+\epsilon_k^{'})&&=\frac{k^{'}}{×kq}\delta\left(\mbox{cos} \theta_{kq}-\frac{\omega}{×q}+\frac{t}
{2kq×}\right).
\eea
 The above delta functions contain terms upto $\omega^2$, though we restrict ourselves upto order $\omega$ in our calculation. Using the above delta functions we obtain,
\bea
\frac{2\mu^4\nu \nu^{'}}{3(2\pi)^5}\int dp \,dk\, f_{\epsilon_p}f_{\epsilon_k} (1- f_{(\epsilon_p+\omega)})
(1- f_{(\epsilon_k-\omega)})dq\, d\omega 
\omega^2(1-x^2)
\int_0^{2\pi}\frac{d\phi}{2\pi}(1-\text{cos}\phi)| M | ^2.
\label{k_num}
\eea
% In presence of the medium the  in the above equation changes. 
The inclusion of the electron self-energy in the dispersion relation changes the phase space factor, which, in turn, changes 
the energy integral in the above equation. We write the momentum integration in the phase-space factor of Eq.(\ref{k_num}) as,
\bea
% d\text{k}=\frac{d\text{k}}{d \epsilon_k}d\epsilon_k
\frac{d\text{k}}{d\epsilon_k}&\simeq& \left(1+\frac{\partial \text{Re}\Sigma}{\partial \omega}\right),
\label{ph_sp}
\eea
where, we have assumed that the quasiparticle energy $\omega$ obeys the following dispersion relation,
\bea
\omega=\epsilon_k-\text{Re}\Sigma(\omega, k).
\label{disp_rel}
\eea
Only the real part of the self-energy has been taken into consideration as the imaginary part turns out to be negligible
 in comparison with the real part. In the free case $d\text{k}/d \epsilon_k$ gives the inverse fermion velocity. 

% Now, to evaluate the remaining integral in the Eq.(\ref{k_num}) one has to specify the form of the matrix amplitude squared.
% \bea
% \frac{ \mu^4T^2}{3(2\pi)^5}\int d\text{q}\, d\omega \frac{\frac{\omega^2}{4T^2}}{\left(\text{sinh}\frac{\omega}{2T}\right)^2}
% \omega^2\left(1-\frac{\omega^2}{q^2}\right)
% \int_0^{2\pi}\frac{d\phi}{2\pi}(1-\text{cos}\phi)| M | ^2
% \eea 
From Eq.(\ref{def_cond}) we can now write the expression for $\kappa_{ee}^{-1}$ as,
\bea
\frac{1}{\kappa_{ee}}=\frac{48 \nu \nu^{'}\alpha^2}{\pi^3 T^2}\frac{d\text{k}}{d\epsilon_k}\frac{d\text{p}}{d\epsilon_p}
I_{\kappa_{ee}}(T/m_D),
\eea
where,
\bea
I_{\kappa_{ee}}(T/m_D)&\equiv&\int_0^{\infty}\frac{d\omega}{\omega}\frac{\frac{\omega^2}{4T^2}}{\left(\text{sinh}\frac{\omega}{2T}\right)^2}
\int_0^1 dx\int_0^{2\pi}\frac{d\phi}{2\pi}x^2(1-x^2)(1-\text{cos}\phi)\nn\
&\times&\Bigg|\frac{1}{1+(xm_D/\omega)^2\chi_L(x)×}-\frac{\text{cos}\phi}{1+(xm_D/\omega)^2\chi_T(x)/(1-x^2)×}\Bigg|^2.
\label{i_kappa}
\eea
 
 In Eq.(\ref{i_kappa}) the major contribution comes from the small angle approximation {\em i.e} 
 the small value of $x$ dominates ($x\sim \omega/ m_D$ or equivalently $q\sim m_D$). We have to consider the small $x$ 
behavior of $\Pi_{L,T}$, we approximate the polarization function in the small $x$ 
($|x|<<1$) limit to obtain, 
\bea
\chi_L&=&1+O(x),\quad \chi_T(x)=i\frac{\pi x}{4}+O(x^2).
\eea 
In the next two subsections we present the results of the thermal relaxation time both for the leading and the higher orders.   
% \tau_{\kappa}=
% \eea
%%%%%%%%%%%%%%%%%%%%%%%%%%%%%%%%%%%%%%%%%%%%%%%%%%%%%%%%%%%%%%%%%%%%%%%%%%%%%%%%%%%%%%%%%%%%%%%%%%%%%%%%%%%%%%%%%%%%%%%
\subsection{Leading order thermal relaxation time}
%%%%%%%%%%%%%%%%%%%%%%%%%%%%%%%%%%%%%%%%%%%%%%%%%%%%%%%%%%%%%%%%%%%%%%%%%%%%%%%%%%%%%%%%%%%%%%%%%%%%%%%%%%%%%%%%%%%%%%%%
In this section we derive the LO result of the thermal relaxation time in the degenerate electron plasma
 present in the outer crust of the neutron star. It might be recalled that in the FL theory the magnetic contribution
 is suppressed compared to the electric one. In this domain $\kappa_{ee}$ varies inversely with $T$ and $c_v\propto T$.
 The thermal relaxation time on the other hand is the ratio of these two quantities. Therefore, in FL theory or in absence
 of transverse interaction $\tau_{\kappa_{ee}}\propto 1/T^2$. Now, we incorporate relativistic effects in it in the small energy transfer region ($\omega<<m_D$, $T<<m_D$)
 where transverse, weak dynamical screening effect becomes important. For this, transverse interaction has been incorporated in the 
thermal conductivity and 
 medium modified phase-space factor has also been included. Now, to have the LO expression of $\tau_{\kappa_{ee}}$ 
we have to evaluate $I_{\kappa_{ee}}(T/m_D)$ in Eq.(\ref{i_kappa}). In this region where $\omega<<m_D$, $T<<m_D$
 the upper
 limit of the $x$ integral can be sent to infinity. Electric interaction in this region gives the following contribution,
\bea
I_L&=&\int_0^{\infty} \frac{d\omega}{\omega}
\frac{\omega^2}{4T^2\left(\text{sinh}\frac{\omega}{2T}\right)^2}\int_0^{\infty}dx
\frac{x^2\left(1-x^2\right)}{\left(1+
\frac{m_D^2x^2}{\omega^2}\right)^2}\nn\
&=&\frac{ \pi ^5 }{15}\lambda^3+
\frac{4 \pi ^7 }{7}\lambda^5.
\label{long_2}
\eea
where, $\lambda$ is the small temperature expansion parameter $T/m_D$. The magnetic interaction on the other hand gives,
\bea
I_T&=&\int_0^{\infty} \frac{d\omega}{\omega}
\frac{\omega^2}{4T^2\left(\text{sinh}\frac{\omega}{2T}\right)^2}\int_0^{\infty}dx
\frac{x^2\left(1-x^2\right)}{1+\frac{\pi^2m_D^4x^6}{16\omega^4}}\nn\
&\simeq&2\lambda^2\zeta(3).
% &=&\int_0^{\infty} \frac{d\omega}{\omega}
%  \frac{\omega^2}{4T^2\left(\text{sinh}\frac{\omega}{2T}\right)^2}\left(-2m_D^{4/3}\omega^2+ \frac{8 (-1)^{2/3} 2^{1/3} \omega ^{10/3}}{3\pi ^{2/3}
% m_D^{10/3}}\right)
\label{trans_2}
\eea   
% Now, the first term gives the following contribution,
% \bea
% -\int_0^{\infty}d\omega
% \frac{m_D^{4/3}\omega^3}{4T^2\left(\text{sinh}\frac{\omega}{2T}\right)^2}=-2\left(\frac{T}{m_D}\right)^2\zeta(3)
% \eea
The higher order contribution ($(T/m_D)^{10/3}$) in the above equation can be neglected.
For the cross term we find that,
\bea
I_{L,T}&=&\int_0^{\infty} \frac{d\omega}{\omega}
\frac{\omega^2}{4T^2\left(\text{sinh}\frac{\omega}{2T}\right)^2}\int_0^{\infty}dx
\frac{x^2\left(1-x^2\right)}{\left(1+
\frac{m_D^2x^2}{\omega^2}\right)\left(1+\frac{\pi^2m_D^4x^6}{16\omega^4}\right)}\nn\
&\simeq&\frac{(2 \pi )^{2/3} }{3 }\lambda^{8/3} \zeta \left(\frac{11}{3}\right)
 \Gamma \left(\frac{14}{3}\right).
\label{cross_2}
\eea
From the above Eqs.(\ref{long_2}, \ref{trans_2}, \ref{cross_2}) we see that the term coming from the magnetic sector 
dominates over the electric and the crossed terms. 
% But the coefficients of these terms make them comparable with the 
%  transverse one and this has been discussed later in this paper.
After the angular integration we now focus on the momentum integral. The momentum integration gives,
% \bea
% \frac{d\text{k}}{d\epsilon_k}&\simeq& \left(1+\frac{\partial \text{Re}\Sigma}{\partial \omega}\right)\nn\
% &=&1+\frac{e^2}{12 \pi ^2} \log \left(\frac{4 m_D}{\pi  (\epsilon -\mu )}\right)
% +\frac{2^{2/3} e^2 
% (\epsilon -\mu )^{2/3}}{9 \sqrt{3} \pi ^{7/3} m_D^{2/3}}-\frac{40\times 2^{1/3} e^2 \left(\epsilon-\mu\right) ^{4/3} }
% {27 \sqrt{3} \pi ^{11/3} m_D^{4/3}}\cdots\nn\
% &=&(1+\beta),
% \eea
\bea
\int dp \,dk\, f_{\epsilon_p}f_{\epsilon_k} (1- f_{(\epsilon_p+\omega)})
(1- f_{(\epsilon_k-\omega)})
% &=&\int  \frac{d\text{p}}{d\epsilon_p}d\epsilon_p\frac{d\text{k}}{d\epsilon_k×}d\epsilon_k
%  f_{\epsilon_p}f_{\epsilon_k} (1\pm f_{(\epsilon_p+\omega)})
% (1\pm f_{(\epsilon_k-\omega)})\nn\
=\left(1+\gamma\right)
T^2 \frac{\frac{\omega^2}{4T^2}}{\left(\text{sinh}\frac{\omega}{2T}\right)^2}
,
\label{nfl_phase}
\eea
where, we have used Eqs.(\ref{ph_sp}, \ref{disp_rel}). The details of this integration is given in the appendix. In the above integral $\gamma$ represents the medium effect in the 
phase space factor. In the LO, $\gamma$ receives the 
logarithmic correction from the quasiparticle self-energy,
\bea
\Sigma(\epsilon, k) &=&e^2m\,
 \Big\{{\epsilon\over12\pi^2m}\Big[\log\Big({4\sqrt{2}m\over\pi \epsilon}\Big)+1\Big] \Big\},
\label{ferm_se}
\eea
where, $\epsilon$ is chosen to be $(\epsilon_k-\mu)$ and in the low temperature limit $(\epsilon_k-\mu)\sim T$. The approximation is sufficiently
 accurate as only a narrow energy level near the Fermi surface is responsible for heat conduction of strongly degenerate 
particles, $m$ is related to the Debye mass ($m_D$) through the relation $m^2=m_D^2/2$. Hence, $\gamma$ can 
 be
 expressed as,
\bea
\gamma=\left[\frac{e^2}{12 \pi ^2 }\log \left(\frac{4  }{\pi  \lambda}\right)\right].
\eea
Thermal electron conductivity now takes the following form,
\bea
 \kappa_{ee}=\Bigg[\left(\frac{48\nu \nu^{'}\left(1+\frac{e^2}{12 \pi ^2 }
\log \left(\frac{4 }{\pi  \lambda}\right)\right)}{\pi^3}\alpha^2T^{-2}\right)\Bigg\{\lambda^2\zeta(3)+\frac{(2 \pi )^{2/3} }{3 }\lambda^{8/3} 
\zeta \left(\frac{11}{3}\right)
 \Gamma \left(\frac{14}{3}\right)
+\frac{ \pi ^5 }{15}\lambda^3\Bigg\}\Bigg]^{-1}.\nonumber\\
\label{thermal_cond_lo}
\eea

It is known that the specific heat capacity is $c_v=\mu^2 T/3$ at the LO. For the thermal relaxation time we now have,
 from Eq.(\ref{relax_time}),
\bea
 \tau_{\kappa_{ee}}=\frac{9}{\mu^2T}\Bigg[\left(\frac{48\nu \nu^{'}\left(1+\frac{e^2}{12 \pi ^2 }
\log \left(\frac{4 }{\pi  \lambda}\right)\right)}{\pi^3}\alpha^2T^{-2}\right)\Bigg\{\lambda^2\zeta(3)+\frac{(2 \pi )^{2/3} }{3 }\lambda^{8/3} 
\zeta \left(\frac{11}{3}\right)
 \Gamma \left(\frac{14}{3}\right)
+\frac{ \pi ^5 }{15}\lambda^3\Bigg\}\Bigg]^{-1}.\nonumber\\
\label{thermal_relax_time_lo}
\eea
In the above equation the dominant first term in the second bracket is from the magnetic sector, the second term is from the longitudinal
 transverse cross term and the third one is from the electric one. This should be contrasted with the earlier results 
what have been
 reported in \cite{Heiselberg93, Shternin06}. In \cite{Heiselberg93, Shternin06} the authors have mentioned that in the low energy transfer region 
($\omega<<m_D$, $T<<m_D$) 
 the thermal conductivity becomes independent of temperature. This happens if only the LO term in $\lambda$ 
in Eq.(\ref{thermal_relax_time_lo}) is considered, but the coefficients of the other NLO two terms
 make them comparable with the first one. Further modification of the result takes place when medium modified dispersion
 relation is included in the phase space factor of $\tau_{\kappa_{ee}}$. 
Inclusion of $\gamma$ changes the temperature dependence significantly. In this context we can comment here that 
$\gamma$ has to be included in the expression 
of $\tau_{\kappa_{ee}}$ since at very low temperature $\alpha^2\log(T)$ term becomes large in comparison to one. We also
 observe here that the Fermi liquid description breaks
 down with the inclusion of the magnetic interaction. 
%%%%%%%%%%%%%%%%%%%%%%%%%%%%%%%%%%%%%%%%%%%%%%%%%%%%%%%%%%%%%%%%%%%%%%%%%%%%%%%%%%%%%%%%%%%%%%%%%%%%%%%%%%%%%%%%%%%%%%%%%%%%
\subsection{Higher orders in the thermal relaxation time}
%%%%%%%%%%%%%%%%%%%%%%%%%%%%%%%%%%%%%%%%%%%%%%%%%%%%%%%%%%%%%%%%%%%%%%%%%%%%%%%%%%%%%%%%%%%%%%%%%%%%%%%%%%%%%%%%%%%%%%%%%%%%
In this subsection we evaluate the higher order correction terms in the low temperature thermal relaxation time. This has its origin in the inclusion of the NLO terms in the 
specific heat capacity and $\gamma$. A convenient starting
point for this would be to extend the calculation of $\gamma$ in Eq.(\ref{nfl_phase}) which relates itself to the
 electron self-energy ($\Sigma$). In case of the quasiparticle momenta close to the Fermi momentum the $\Sigma$ is 
dominated by the soft photon exchange. Beyond the leading order at low temperature, it is given by the following equation \cite{Gerhold05},
\bea
\Sigma(\epsilon, k) &=&e^2m\,
 \Big\{{\epsilon\over12\pi^2m}\Big[\log\Big({4\sqrt{2}m\over\pi \epsilon}\Big)+1\Big]+{i \epsilon\over24\pi m}
  \,+{2^{1/3}\sqrt{3}\over45\pi^{7/3}}\left({\epsilon\over m}\right)^{5/3}(\mathrm{sgn}(\epsilon)-\sqrt{3}i)\qquad\nn\
  &&+ {i\over64 \sqrt{2}}\left({\epsilon\over m}\right)^2
  -20{2^{2/3}\sqrt{3}\over189\pi^{11/3}}\left({\epsilon\over m}\right)^{7/3}(\mathrm{sgn}(\epsilon)+\sqrt{3}i)\qquad\nn\
&&-{6144-256\pi^2+36\pi^4-9\pi^6\over864\pi^6}\Big({\epsilon\over m} 
  \Big)^3 \Big[\log\left({{0.928}\,m\over \epsilon}\right) %-0.4214
-{i\pi\mathrm{sgn}(\epsilon)\over 2}  \Big]
  +\mathcal{O}\Big(\left({\epsilon\over m}\right)^{11/3}\Big) \Big\}.
\label{ferm_se}
\eea
 The dominant 
logarithmic
 term in the fermion self-energy comes from the transverse sector and it gives rise to logarithmic 
singularity when $\epsilon_k\rightarrow \mu$ or in other words for excitations near the Fermi surface.
We approximate the above expression in the low temperature limit as $(\epsilon_k-\mu)\sim T$, as has been 
done in the last subsection.  With the above quasiparticle self-energy we have,
\bea
\frac{d\text{k}}{d\epsilon_k}&\simeq& \left(1+\frac{\partial \text{Re}\Sigma}{\partial \omega}\right)\nn\
&=&1+\frac{e^2}{12 \pi ^2} \log \left(\frac{4 }{\pi  \lambda}\right)
+\frac{2^{2/3} e^2 
\lambda^{2/3}}{9 \sqrt{3} \pi ^{7/3} }-\frac{40\times 2^{1/3} e^2 \lambda ^{4/3} }
{27 \sqrt{3} \pi ^{11/3}}\cdots\nn\
&=&(1+\beta),
\eea

%  The details of the energy integral has been evaluated in the appendix. Now, we first see how the phase-space integral can be evaluated,
% \bea
% \int d\text{p} \,d\text{k}\, f_{\epsilon_p}f_{\epsilon_k} (1\pm f_{(\epsilon_p+\omega)})
% (1\pm f_{(\epsilon_k-\omega)})
% &=&\int  \frac{d\text{p}}{d\epsilon_p}d\epsilon_p\frac{d\text{k}}{d\epsilon_k×}d\epsilon_k
%  f_{\epsilon_p}f_{\epsilon_k} (1\pm f_{(\epsilon_p+\omega)})
% (1\pm f_{(\epsilon_k-\omega)})\nn\
% =\left(1+\gamma\right)
% T^2 \frac{\frac{\omega}{4T^2}}{\left(\text{sinh}\frac{\omega}{2T}\right)^2}

% \label{nfl_phase}
% \eea
where, $\beta=\gamma+\gamma^{'}$. $\gamma^{'}$ gives us the NLO NFL terms,
\bea
\gamma^{'}=\frac{2^{2/3} e^2 
\lambda^{2/3}}{9 \sqrt{3} \pi ^{7/3} }-\frac{40\times 2^{1/3} e^2 \lambda ^{4/3} }
{27 \sqrt{3} \pi ^{11/3} }.
\eea
%  and this brings the major departure of the present result from the previous one.
The final expression for the electron thermal conductivity now becomes,
\bea
\kappa_{ee}&=&\Bigg[\frac{48\nu\nu^{'}}{\pi^3}\alpha^2T^{-2}\left(1+\beta\right)\Big\{\lambda^2\zeta(3)+
\frac{(2 \pi )^{2/3} }{3 }\lambda^{8/3} 
\zeta \left(\frac{11}{3}\right)
 \Gamma \left(\frac{14}{3}\right)\nn\
&+&\frac{ \pi ^5 }{15}\lambda^3\Big\}\Bigg]^{-1}.
\label{cond_final_2}
\eea
 Unlike Fermi-liquid result where, 
$\kappa_{ee}$ varies inversely with $T$, here the temperature dependence is non-analytical, anomalous in nature reminiscent
 of many other recent studies involving ultradegenerate plasma \cite{Gerhold05, Sarkar11, Pal11, Adhya12}. We show here that $\kappa_{ee}$ involves fractional power in $(T/m_D)$ 
coming from the medium modified phase space factor. 
% The
%  reason behind these anomalous terms 
% are from the weak dynamical screening in the magnetic sector. 

The other quantity which we require for the estimation of relaxation time is the specific heat. It has already been derived in the 
context of degenerate quark matter in \cite{Ipp04, Gerhold204}. For the degenerate electron gas it can be written as,
\bea
\label{spec-heat}
c_v&=&\frac{\mu^2T}{3}+{m_D^2 T\over 36}\left(\ln\left({4\over\pi \lambda}\right)+\gamma_E
  -{6\over\pi^2}\zeta^\prime(2)-3\right)\nonumber\\
  &&-40{2^{2/3}\Gamma\left({8\over3}\right)\zeta\left({8\over3}\right)m_D^3\over27\sqrt{3}\pi^{7/3}}
  \lambda^{5/3}
  +560{2^{1/3}\Gamma\left({10\over3}\right)\zeta\left({10\over3}\right)m_D^3
  \over81\sqrt{3}\pi^{11/3}}\lambda^{7/3}.\nonumber\\
\label{sp_heat}
 \eea
With Eqs.(\ref{cond_final_2}) and (\ref{spec-heat}) the relaxation time for thermal conduction is found to be,

% \bea
\bea
 \tau_{\kappa_{ee}}&=&3\Bigg[\left(\frac{48\nu \nu^{'}\left(1+\beta\right)}{\pi^3}\alpha^2T^{-2}\right)
\Bigg\{\lambda^2\zeta(3)+\frac{(2 \pi )^{2/3} }{3 }\lambda^{8/3} 
\zeta \left(\frac{11}{3}\right)
 \Gamma \left(\frac{14}{3}\right)
\nn\
&+&\frac{ \pi ^5 }{15}\lambda^3\Bigg\}\Bigg]^{-1}\Bigg/\Bigg[\frac{\mu^2T}{3}+{m_D^2 T\over 36}\left(\ln\left({4\over\pi \lambda}\right)+\gamma_E
  -{6\over\pi^2}\zeta^\prime(2)-3\right)\nonumber\\
  &&-40{2^{2/3}\Gamma\left({8\over3}\right)\zeta\left({8\over3}\right)m_D^3\over27\sqrt{3}\pi^{7/3}}
  \lambda^{5/3}
  +560{2^{1/3}\Gamma\left({10\over3}\right)\zeta\left({10\over3}\right)m_D^3
  \over81\sqrt{3}\pi^{11/3}}\lambda^{7/3}\Bigg].\nonumber\\
\label{thermal_relax_time}
\eea
It is evident that the thermal relaxation time upto NLO terms contains some anomalous
 fractional powers originated from the transverse interaction. This in turn changes the temperature dependence of 
$\tau_{\kappa_{ee}}$ non-trivially. The appearance of the non-analytic 
terms in Eqs.(\ref{cond_final_2}), (\ref{spec-heat}) and (\ref{thermal_relax_time}) has a common origin as explained earlier.

%%%%%%%%%%%%%%%%%%%%%%%%%%%%%%%%%%%%%%%%%%%%%%%%%%%%%%%%%%%%%%%%%%%%%%%%%%%%%%%%%%%%%%%%%%%%%%%%%%%%%%%%%%%%%%%%%%%%%%%%%%%
\section{Results and Discussion}
%%%%%%%%%%%%%%%%%%%%%%%%%%%%%%%%%%%%%%%%%%%%%%%%%%%%%%%%%%%%%%%%%%%%%%%%%%%%%%%%%%%%%%%%%%%%%%%%%%%%%%%%%%%%%%%%%%%%%%%%%%%
\medskip
\begin{figure}[htb]
\begin{center}
\resizebox{8.5cm}{5.75cm}{\includegraphics{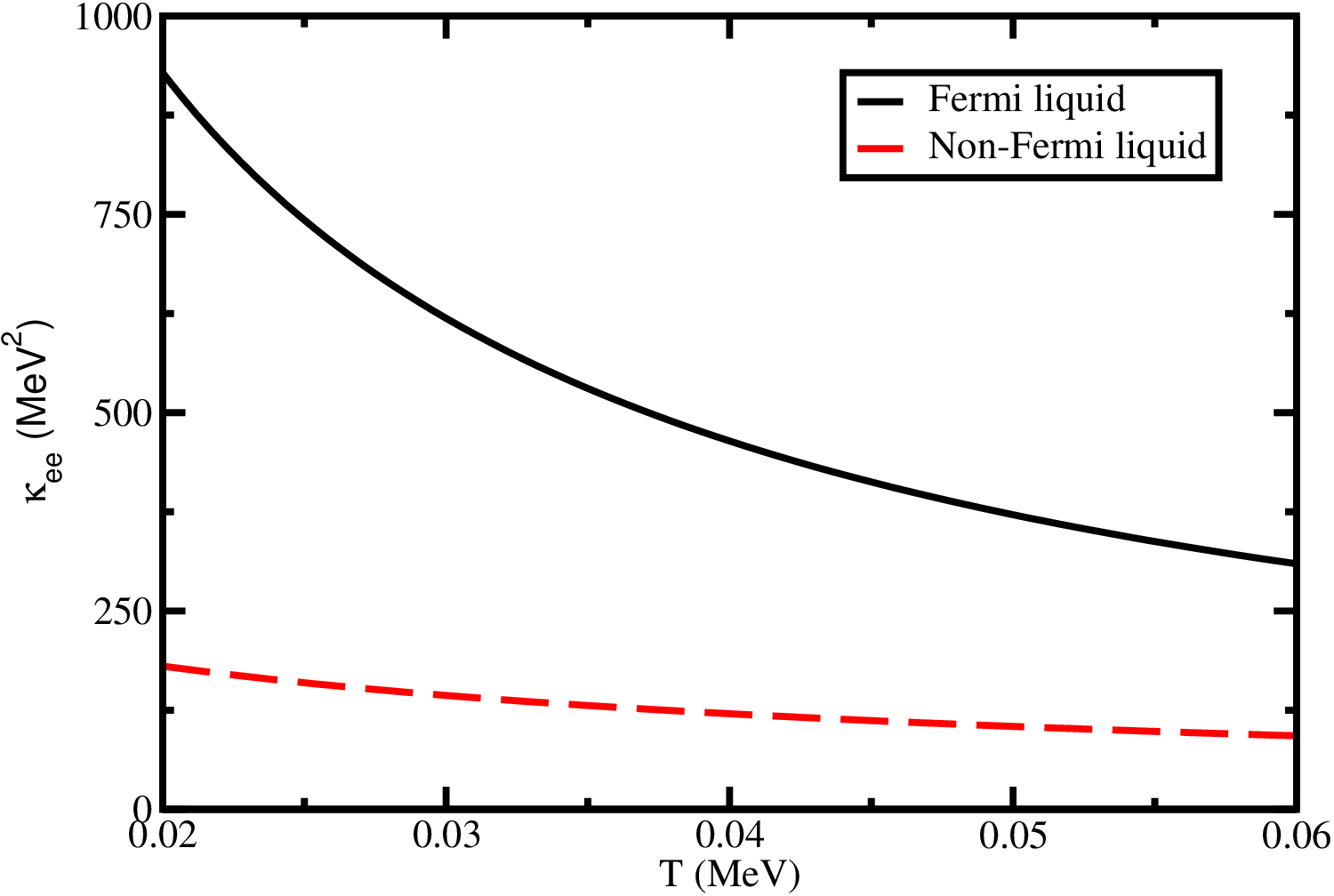}}
~~~~~~~~\resizebox{8.5cm}{5.75cm}{\includegraphics{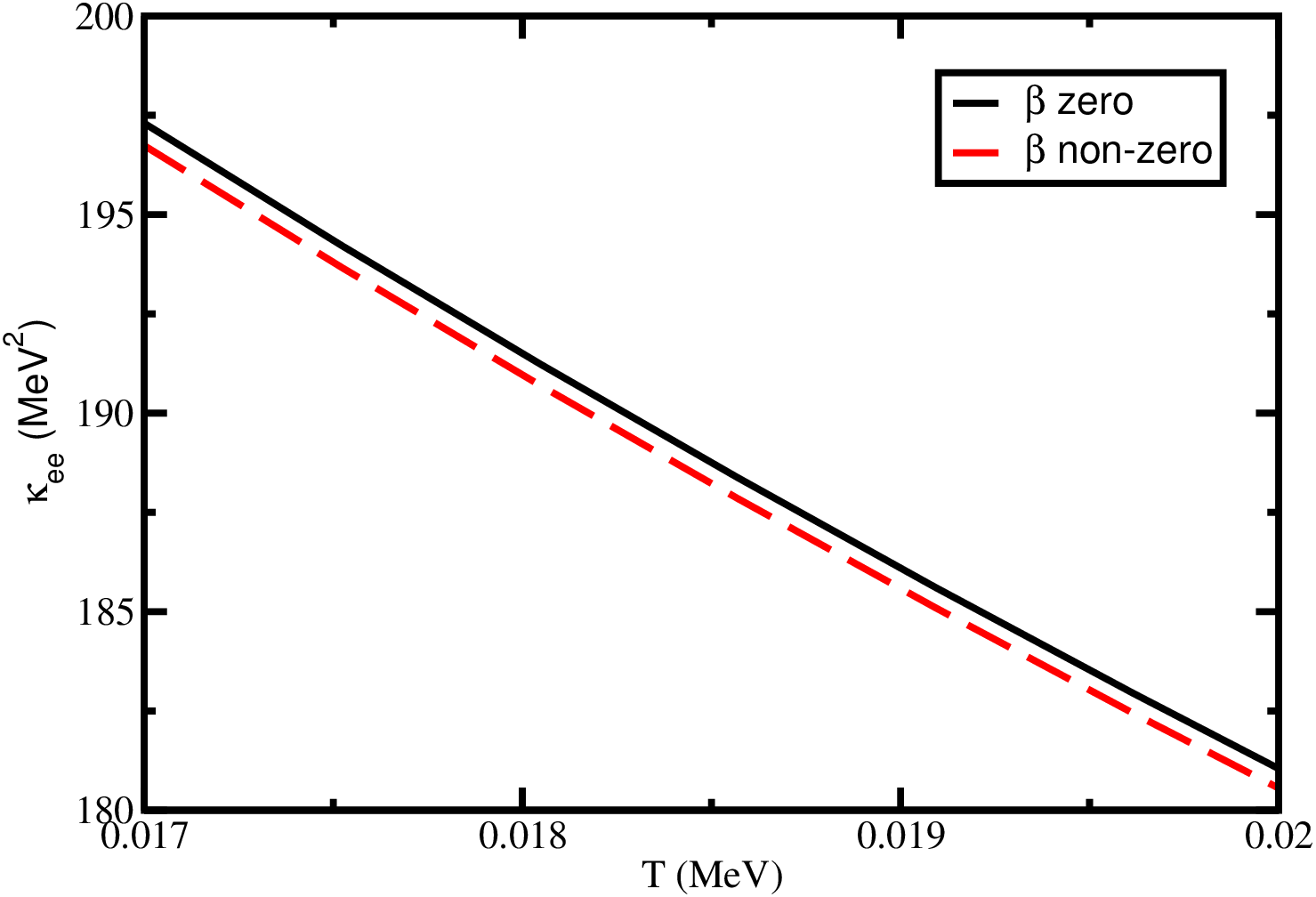}}
\caption{Temperature dependence of the electron thermal conductivity. The left panel shows a comparison between the 
Fermi liquid and the non-Fermi-liquid NLO result where $\beta \neq 0$. The right panel shows the reduction of $\kappa_{ee}$
 with the inclusion of $\beta$.
% solid line corresponds
%  to the case of the thermal conductivity without phase-space modification and leading order
%  specific heat. The dot-and-dashed line includes both the phase-space modified $\kappa_{ee}$ and the specific heat with NLO correction.
\label{fig1}}
\end{center}
\end{figure}
% \bigskip
% \medskip

In this section an estimation of the electron thermal conductivity and the electron thermal relaxation time with the
 temperature has been presented. In Fig.(\ref{fig1}) we have plotted $\kappa_{ee}$ with $T$ using  
 Eq.(\ref{cond_final_2}). In the left panel of Fig.(\ref{fig1}) we note that the inclusion
 of both the medium modified propagator and $\beta$ decreases the value of $\kappa_{ee}$. It shows strong deviation from the 
Fermi liquid result $\kappa_{ee}\propto 1/T$. In the right panel it has been shown that $\beta$ reduces thermal conductivity. 
This has serious implication on the total electron conductivity $\kappa_e$. In \cite{Shternin06} the
 authors have shown that magnetic interaction decreases $\kappa_{ee}$ which in turn increases the electron-electron collision
 frequency. Thus to the total electron thermal conductivity electron-electron scattering dominates over electron-ion
 scattering. The phase space correction due to the medium modification of the electron dispersion relation further enhances the
  electron-electron collision frequency.
 
%  The phase space correction due to the medium modification of the electron dispersion relation partially 
% compensates for such an enhancement.
% Due to this the contribution of $\kappa_{ee}$ in total electron thermal conductivity increases in comparison with
%  $\kappa_{ei}$ one. But, here we have observed that phase-space modification causes enhancement in $\kappa_{ee}$.

\begin{figure}[h]
\begin{center}
\resizebox{8.5cm}{5.75cm}{\includegraphics{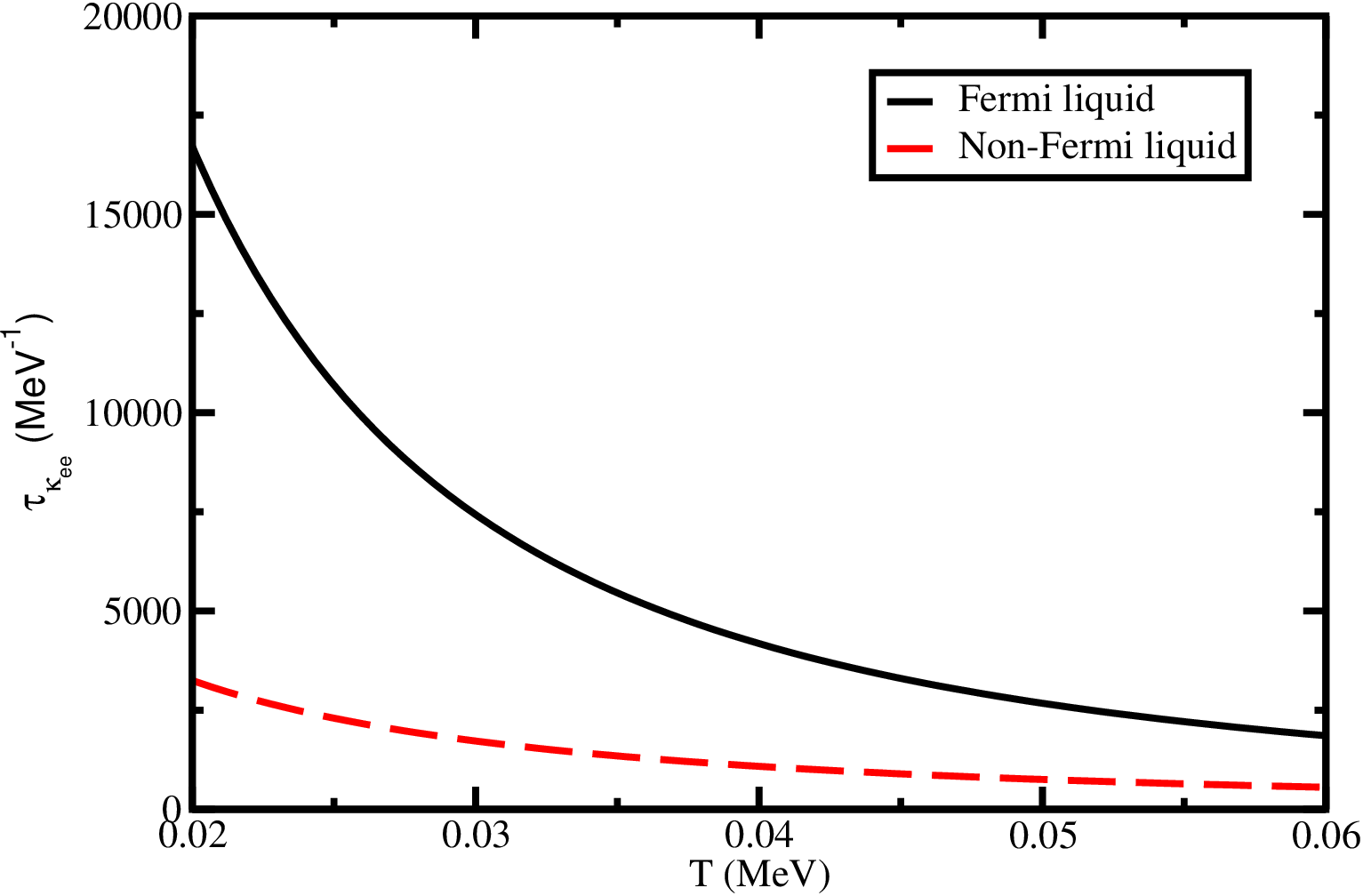}}
% \caption{Temperature dependence of the thermal relaxation time. The solid line corresponds
%  to the case of the thermal conductivity without phase-space modification and leading order
%  specific heat. The dot-and-dashed line includes both the phase-space modified $\kappa_{ee}$ and the specific heat with NLO correction.
% \begin{figure}[h]
% \begin{center}
\resizebox{8.5cm}{5.75cm}{\includegraphics{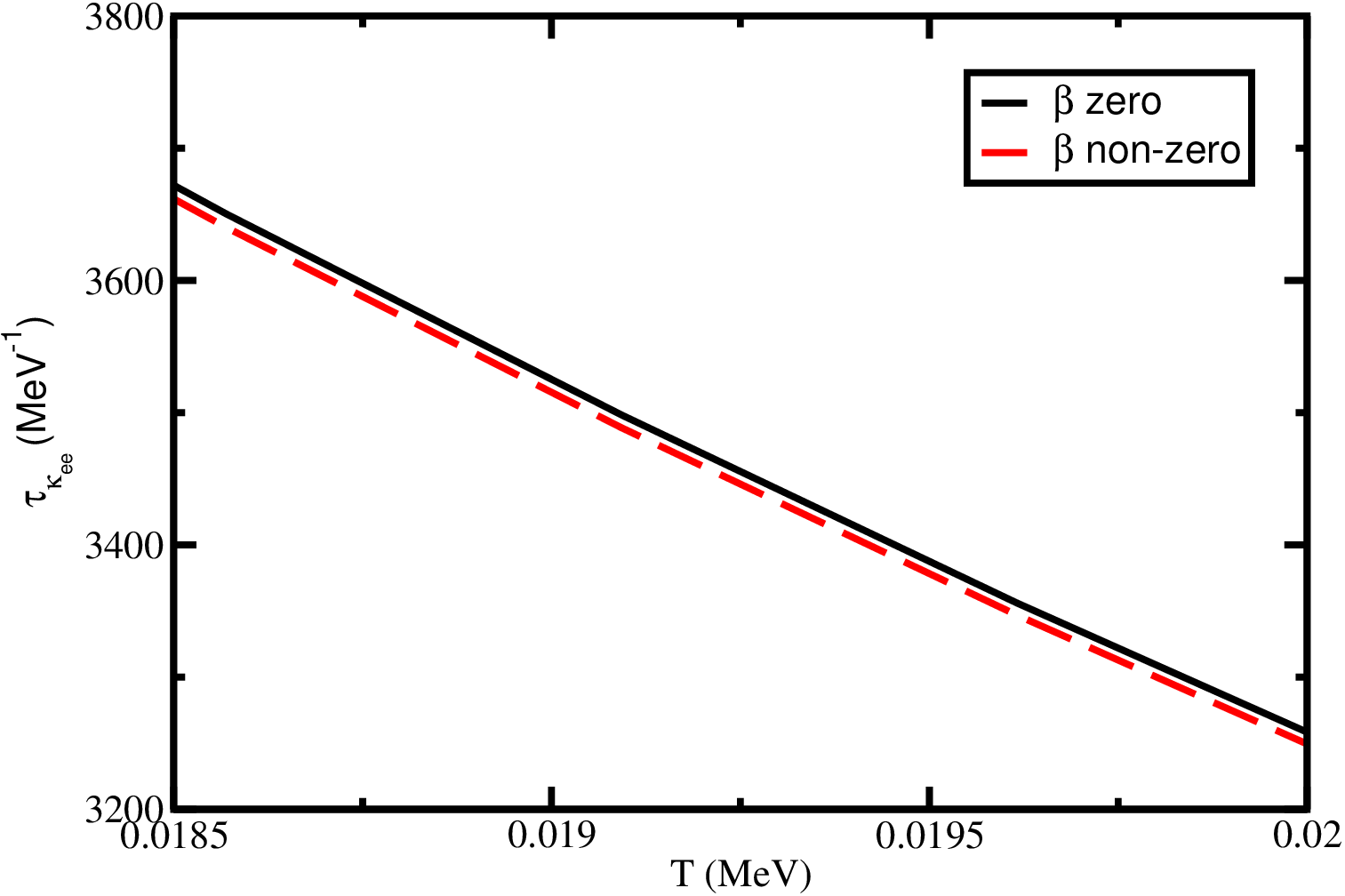}}
\caption{Temperature dependence of the thermal relaxation time. The left panel shows a comparison between the 
Fermi liquid and the non-Fermi liquid NLO result when $\beta\neq 0$. The right panel shows the reduction of $\tau_{\kappa_{ee}}$
 with the inclusion of $\beta$.
% The solid line corresponds
%  to the case of the thermal conductivity without phase-space modification and leading order
%  specific heat. The dot-and-dashed line includes both the phase-space modified $\kappa_{ee}$ and the specific heat with NLO correction.
% \label{fig2}}
\label{fig3}}
\end{center}
\end{figure}

%  In Fig.(\ref{fig2}) we have shown the temperature
%  dependence of the specific heat capacity of the electron matter. It reveals the reduction of $c_v$ with the NFL correction. 
 With the thermal conductivity
 and specific heat capacity we further plot the thermal relaxation time with the temperature using Eq.(\ref{thermal_relax_time}). In Fig.(\ref{fig3}) 
we have shown how the thermal
 relaxation time changes from the FL result with the inclusion of the medium modifications. In the right panel it has been
 shown that the inclusion of $\beta$ reduces the thermal relaxation time.
% From the plots it is clear that medium modification
% %  decreases the value of thermal relaxation time in comparison with the FL result. This reduction has its significant role
%  in the time of heat diffusion from the inner crust to the surface.
 We have taken the relevant region of temperature
  and density from the reference \cite{Shternin06}. According to \cite{Shternin06} in the density region higher than $\rho=10^6$ 
$\text{gcm}^{-1}$ and temperature more than $10^8$ K Landau damping becomes important since in this density region 
non-relativistic degenerate electron plasma becomes relativistic. This region is relevant for our plots. 
% important We have taken Debye mass $344$ KeV, this satisfies the condition of $T<<m_D$ where medium modification becomes important.

% \begin{figure}[htb]
% \begin{center}
% \resizebox{8.5cm}{5.75cm}{\includegraphics{sp_ht.eps}}
% \caption{Temperature dependence of the specific heat. 
% \label{fig2}}
% \end{center}
% \end{figure}
%  \bigskip
% \medskip

 \bigskip
% \medskip
%%%%%%%%%%%%%%%%%%%%%%%%%%%%%%%%%%%%%%%%%%%%%%%%%%%%%%%%%%%%%%%%%%%%%%%%%%%%%%%%%%%%%%%%%%%%%%%%%%%%%%%%%%%%%%%%%%%%%%
\section{Summary}
%%%%%%%%%%%%%%%%%%%%%%%%%%%%%%%%%%%%%%%%%%%%%%%%%%%%%%%%%%%%%%%%%%%%%%%%%%%%%%%%%%%%%%%%%%%%%%%%%%%%%%%%%%%%%%%%%%%%%%
In this paper, we calculate the thermal relaxation time in degenerate electron gas in the domain where the relativistic
 effects become important. It has been shown that with the inclusion of the magnetic interaction, which is relevant
 in the above mentioned domain, $\tau_{\kappa_{ee}}$ shows departure from the FL behavior. It is known that for the normal
 FL {\em i.e} with only Coulomb interaction $\tau_{\kappa_{ee}}$ behaves as $1/T^2$. In the relativistic domain, we on the other
 hand, find this
 to be proportional to the inverse temperature {\em i.e} $\tau_{\kappa_{ee}}\propto 1/T$. This has been attributed
 to the absence of screening in the transverse sector. We further expose how the in medium modifications of 
the electron dispersion characteristic affects the 
heat conduction from the neutron star crust to the core. Our calculation actually modifies the phase space or the Fermionic density
 of states as revealed in the text leading to a reduction of conductivity or relaxation time. For the thermal
 relaxation time, a closed form analytical expression has been derived by making perturbative expansion in $(T/m_D)$ 
retaining terms beyond LO. The appearance
 of the fractional powers in these results are interesting which is reminiscent of what one obtains in the 
calculation of the fermionic self-energy at high density. Numerically these corrections have also been found to be important
 in the present context. Particularly these corrections become important in the domain of small frequency {\em i.e} when $\omega<<
m_D$.
% In this paper we have shown that inclusion of the magnetic interaction in the thermal relaxation time in degenerate electron gas 
% changes its temperature dependence from $1/T^2$ to $1/T$ in the neutron star crust. 
%  The other important aspect 
% of the present work has been the inclusion of the magnetic interaction in the calculation of $\tau_{\kappa_{ee}}$ where
%  a perturbative expansion of this quantity has been made in   These corrections have also been found to be numerically important
%  in the present context.  {\bf Furthermore, we show that with the magnetic interaction and the phase-space modification electron-electron
%  scattering wins over the electron-ion scattering leading to a reduced $\kappa_{ee}$ or $\tau_{\kappa_{ee}}$. In the appropriate
%  limit our results for both of these quantities agree with the previous results.}
\section*{Acknowledgement}
S. Sarkar would like to acknowledge helpful discussions with R. Nandi and S. P. Adhya.

\appendix
\section*{Appendix}
\label{appendixBj}
In this appendix we provide the complete details of the derivation of the energy integral in the phase space factor discussed
 in the main text. We substitute $-(\epsilon_p-\mu)/T=\eta_p$, $-(\epsilon_k-\mu)/T=\eta_k$ and $-\omega/T=z$ to obtain,
\bea
\int_0^{\infty}\, f_{\epsilon_p}f_{\epsilon_k} (1\pm f_{(\epsilon_p+\omega)})
(1\pm f_{(\epsilon_k-\omega)}) d\epsilon_p \,d\epsilon_k=T^2\int_0^{\infty}\frac{1}{\left(\text{e}^{-\eta_p}+1\right)
\left(\text{e}^{(\eta_p+z)}+1\right) 
\left(\text{e}^{-\eta_k}
+1\right)\left(\text{e}^{(\eta_k-z)}+1\right)}d\eta_p d\eta_k\,.
% &=&T\frac{\Delta}{1-\text{e}^{-\Delta}}
\eea
The lower integration limit can be send to infinity without introducing much error. Now, following \cite{Wilson, Zeman} the general way to calculate the integral (for a function $f=\eta^s$, where s is an integer) 
as follows,
\bea
I_s=\int_{-\infty}^{\infty}\frac{\eta^s}{×(\text{e}^{-\eta}+1)(\text{e}^{\eta+z})}d\eta
=\frac{2}{(1+s)(\text{e}^z-1)×}\sum_{r=0,1,\cdots}^{[\frac{1}{2}s]}(-1)^{s-2r} \dbinom{s+1}{2r}  \frac{(s+1)!}{2r!(s+1-2r)!}(2r)!
\mathrm{C}_{2r}z^{s+1-2r},
\eea
where,
\bea
2\mathrm{C}_{2r}&=&\frac{1}{2r!}\int_{-\infty}^{\infty}\eta^{2r}\text{e}^{-\eta}
\sum_0^{\infty}(1+s)(-1)^s\text{e}^{-\eta s}d\eta\nn\
&=& 2\sum_0^{\infty}\frac{(-1)^s}{(1+s)^{2r}}.
\eea
 $[\frac{1}{2}s]$ is the integral part of $\frac{1}{2}s$ and $C_0=\frac{1}{2}$. Hence, one can write,
\bea
T\int_0^{\infty}\frac{1}{\left(\text{e}^{-\eta_p}+1\right)\left(\text{e}^{(\eta_p+z)}+1\right)}\, d\eta_p=T\frac{z}{1-\text{e}^{-z}}.
\eea
Finally, we obtain the result of the phase space energy integration quoted in Eq.(\ref{nfl_phase}) in the main text,
\bea
\int d\epsilon_p \,d\epsilon_k\, f_{\epsilon_p}f_{\epsilon_k} (1\pm f_{(\epsilon_p+\omega)})
(1\pm f_{(\epsilon_k-\omega)})
=T^2 \frac{\frac{\omega^2}{4T^2}}{\left(\text{sinh}\frac{\omega}{2T}\right)^2}.
\eea

%%%%%%%%%%%%%%%%%%%%%%%%%%%%%%%%%%%%%%%%%%%%%%%%%%%%%%%%%%%%%%%%%%%%%%%%%%%%%%%
\end{document}